# Magnetic-field Order in the Southwestern Rim of RCW 86 Constrained Using X-Ray Polarimetry

Stefano Silvestri[1], Dmitry Prokhorov[2], Jacco Vink[3], Patrick Slane[4], Yi-Jung Yang[5,6], Niccolò Bucciantini[7,8,9], Riccardo Ferrazzoli[10], Ping Zhou[11], Enrico Costa[10], Nicola Omodei[12], Chi-Yung Ng[13], Paolo Soffitta[10], Martin C. Weisskopf[14], Luca Baldini[1,15], Alessandro Di Marco[10], Victor Doroshenko[16], Jeremy Heyl[17], Philip Kaaret[14], Frédéric Marin[18], Tsunefumi Mizuno[19], Melissa Pesce-Rollins[1], Carmelo Sgrò[1], Douglas A. Swartz[20], Toru Tamagawa[21], Fei Xie[10,22], Iván Agudo[23], Lucio A. Antonelli[24,25], Matteo Bachetti[26], Wayne H. Baumgartner[27], Ronaldo Bellazzini[1], Stefano Bianchi[28], Stephen D. Bongiorno[14], Raffaella Bonino[29,30], Alessandro Brez[1], Fiamma Capitanio[10], Simone Castellano[1], Elisabetta Cavazzuti[31], Chien-Ting Chen[20], Stefano Ciprini[25,32], Alessandra De Rosa[10], Ettore Del Monte[10], Laura Di Gesu[31], Niccolò Di Lalla[12], Immacolata Donnarumma[31], Michal Dovčiak[33], Steven R. Ehlert[14], Teruaki Enoto[21], Yuri Evangelista[10], Sergio Fabiani[10], Javier A. García[34], Shuichi Gunji[35], Kiyoshi Hayashida[36,56], Wataru Iwakiri[37], Svetlana G. Jorstad[38,39], Vladimir Karas[33], Fabian Kislat[40], Takao Kitaguchi[21], Jeffery J. Kolodziejczak[14], Henric Krawczynski[41], Fabio La Monaca[10,42,43], Luca Latronico[29], Ioannis Liodakis[44], Simone Maldera[29], Alberto Manfreda[45], Andrea Marinucci[31], Alan P. Marscher[38], Herman L. Marshall[46], Francesco Massaro[29,30], Giorgio Matt[28], Ikuyuki Mitsuishi[47], Fabio Muleri[10], Michela Negro[48], Stephen L. O'Dell[14], Chiara Oppedisano[29], Alessandro Papitto[24], George G. Pavlov[49], Abel L. Peirson[12], Matteo Perri[24,25], Pierre-Olivier Petrucci[50], Maura Pilia[26], Andrea Possenti[26], Juri Poutanen[51], Simonetta Puccetti[25], Brian D. Ramsey[14], John Rankin[52], Ajay Ratheesh[10,53], Oliver J. Roberts[20], Roger W. Romani[12], Gloria Spandre[1], Fabrizio Tavecchio[52], Roberto Taverna[54], Yuzuru Tawara[47], Allyn F. Tennant[14], Nicholas E. Thomas[14], Francesco Tombesi[32,42], Alessio Trois[26], Sergey S. Tsygankov[51], Roberto Turolla[54,55], Kinwah Wu[55], and Silvia Zane[55]

[1] Istituto Nazionale di Fisica Nucleare, Sezione di Pisa, Largo B. Pontecorvo 3, 56127 Pisa, Italy; stefano.silvestri@pi.infn.it
[2] Fakultät für Physik und Astronomie, Julius-Maximilians-Universität Würzburg, Emil-Fischer-Str. 31, 97074, Würzburg, Germany
[3] Anton Pannekoek Institute for Astronomy & GRAPPA, University of Amsterdam, Science Park 904, 1098 XH Amsterdam, The Netherlands
[4] Center for Astrophysics, Harvard & Smithsonian, 60 Garden St, Cambridge, MA 02138, USA
[5] Center for Astrophysics and Space Science (CASS), New York University Abu Dhabi, PO Box 129188, Abu Dhabi, UAE
[6] Laboratory for Space Research, The University of Hong Kong, Cyberport 4, Hong Kong, People's Republic of China
[7] INAF Osservatorio Astrofisico di Arcetri, Largo Enrico Fermi 5, 50125 Firenze, Italy
[8] Dipartimento di Fisica e Astronomia, Università degli Studi di Firenze, Via Sansone 1, 50019, Sesto Fiorentino (FI), Italy
[9] Istituto Nazionale di Fisica Nucleare, Sezione di Firenze, Via Sansone 1, 50019, Sesto Fiorentino (FI), Italy
[10] INAF Istituto di Astrofisica e Planetologia Spaziali, Via del Fosso del Cavaliere 100, 00133 Roma, Italy
[11] School of Astronomy and Space Science, Nanjing University, Nanjing 210023, People's Republic of China
[12] Department of Physics and Kavli Institute for Particle Astrophysics and Cosmology, Stanford University, Stanford, CA 94305, USA
[13] Department of Physics, The University of Hong Kong, Pokfulam, Hong Kong, People's Republic of China
[14] NASA Marshall Space Flight Center, Huntsville, AL 35812, USA
[15] Dipartimento di Fisica, Università di Pisa, Largo B. Pontecorvo 3, 56127 Pisa, Italy
[16] Institut für Astronomie und Astrophysik, Universität Tübingen, Sand 1, 72076, Tübingen, Germany
[17] University of British Columbia, Vancouver, BC V6T 1Z4, Canada
[18] Université de Strasbourg, CNRS, Observatoire Astronomique de Strasbourg, UMR 7550, 67000 Strasbourg, France
[19] Hiroshima Astrophysical Science Center, Hiroshima University, 1-3-1 Kagamiyama, Higashi-Hiroshima, Hiroshima 739-8526, Japan
[20] Science and Technology Institute, Universities Space Research Association, Huntsville, AL 35805, USA
[21] RIKEN Cluster for Pioneering Research, 2-1 Hirosawa, Wako, Saitama 351-0198, Japan
[22] Guangxi Key Laboratory for Relativistic Astrophysics, School of Physical Science and Technology, Guangxi University, Nanning 530004, People's Republic of China
[23] Instituto de Astrofísica de Andalucía – CSIC, Glorieta de la Astronomía s/n, 18008 Granada, Spain
[24] INAF Osservatorio Astronomico di Roma, Via Frascati 33, 00078, Monte Porzio Catone (RM), Italy
[25] Space Science Data Center, Agenzia Spaziale Italiana, Via del Politecnico snc, 00133 Roma, Italy
[26] INAF Osservatorio Astronomico di Cagliari, Via della Scienza 5, 09047, Selargius (CA), Italy
[27] Naval Research Laboratory, 4555 Overlook Ave. SW, Washington, DC 20375, USA
[28] Dipartimento di Matematica e Fisica, Università degli Studi Roma Tre, Via della Vasca Navale 84, 00146 Roma, Italy
[29] Istituto Nazionale di Fisica Nucleare, Sezione di Torino, Via Pietro Giuria 1, 10125 Torino, Italy
[30] Dipartimento di Fisica, Università degli Studi di Torino, Via Pietro Giuria 1, 10125 Torino, Italy
[31] Agenzia Spaziale Italiana, Via del Politecnico snc, 00133 Roma, Italy
[32] Istituto Nazionale di Fisica Nucleare, Sezione di Roma "Tor Vergata," Via della Ricerca Scientifica 1, 00133 Roma, Italy
[33] Astronomical Institute of the Czech Academy of Sciences, Boční II 1401/1, 14100, Praha 4, Czechia
[34] X-ray Astrophysics Laboratory, NASA Goddard Space Flight Center, Greenbelt, MD 20771, USA
[35] Yamagata University, 1-4-12 Kojirakawa-machi, Yamagata-shi 990-8560, Japan
[36] Osaka University, 1-1 Yamadaoka, Suita, Osaka 565-0871, Japan
[37] International Center for Hadron Astrophysics, Chiba University, Chiba 263-8522, Japan
[38] Institute for Astrophysical Research, Boston University, 725 Commonwealth Avenue, Boston, MA 02215, USA
[39] Department of Astrophysics, St. Petersburg State University, Universitetsky pr. 28, Petrodvoretz, 198504, St. Petersburg, Russia
[40] Department of Physics and Astronomy and Space Science Center, University of New Hampshire, Durham, NH 03824, USA
[41] Physics Department and McDonnell Center for the Space Sciences, Washington University in St. Louis, St. Louis, MO 63130, USA
[42] Dipartimento di Fisica, Università degli Studi di Roma "Tor Vergata", Via della Ricerca Scientifica 1, 00133 Roma, Italy
[43] Dipartimento di Fisica, Università degli Studi di Roma "La Sapienza", Piazzale Aldo Moro 5, 00185 Roma, Italy
[44] Institute of Astrophysics, Foundation for Research and Technology - Hellas, Voutes, 70013 Heraklion, Greece

[45] Istituto Nazionale di Fisica Nucleare, Sezione di Napoli, Strada Comunale Cinthia, 80126 Napoli, Italy
[46] MIT Kavli Institute for Astrophysics and Space Research, Massachusetts Institute of Technology, 77 Massachusetts Avenue, Cambridge, MA 02139, USA
[47] Graduate School of Science, Division of Particle and Astrophysical Science, Nagoya University, Furo-cho, Chikusa-ku, Nagoya, Aichi 464-8602, Japan
[48] Department of Physics and Astronomy, Louisiana State University, Baton Rouge, LA 70803, USA
[49] Department of Astronomy and Astrophysics, Pennsylvania State University, University Park, PA 16801, USA
[50] Université Grenoble Alpes, CNRS, IPAG, 38000 Grenoble, France
[51] Department of Physics and Astronomy, 20014, University of Turku, Finland
[52] INAF Osservatorio Astronomico di Brera, Via E. Bianchi 46, 23807, Merate (LC), Italy
[53] Physical Research Laboratory, Thaltej, Ahmedabad, Gujarat 380009, India
[54] Dipartimento di Fisica e Astronomia, Università degli Studi di Padova, Via Marzolo 8, 35131 Padova, Italy
[55] Mullard Space Science Laboratory, University College London, Holmbury St Mary, Dorking, Surrey RH5 6NT, UK

## Abstract

RCW 86 is a supernova remnant whose origin has recently been linked to an off-center explosion within a cavity created by its progenitor star. In the southwestern region, the forward shock is thought to have reached the cavity wall, encountering diverse environmental conditions. We report on the spatially resolved X-ray polarimetric observation of RCW 86 with the Imaging X-ray Polarimetry Explorer (IXPE). In the 2–4.5 keV energy band we find no significant detection of polarization. Employing a dedicated background subtraction procedure and Bayesian spectropolarimetric fitting, we derive 99% upper limits on the polarization degree of the synchrotron component: 15% in higher-statistics regions and 30%–40% in lower-statistics regions. These upper limits on the polarization degree in several regions exclude the possibility of a strongly coherent magnetic field down to the subparsec scale, and that of a moderately coherent one on the scale of the synchrotron features as resolved by IXPE. The results indicate that the shocks in the southwestern rim of RCW 86 propagate more slowly than the unshocked ejecta at their locations, yet exceed the measured proper motion speeds. This behavior is consistent with reflected shocks occurring in tenuous regions of the shocked ejecta, distinct from regions that are radio-bright.

*Unified Astronomy Thesaurus concepts:* Polarimetry (1278); Spectropolarimetry (1973); Supernova remnants (1667); Shocks (2086); X-ray astronomy (1810)

## 1. Introduction

RCW 86 (also known as MSH 14-63, or G315.4-2.3) is a supernova remnant (SNR) of debated origin and large angular size (40′, S. Broersen et al. 2014) located not far off the Galactic plane that has been studied across a wide energy band. Optical observations (K. S. Long & W. P. Blair 1990; M. Rosado et al. 1996) have been used to estimate its kinematic distance and expansion speed using the width of the Balmer filaments encompassing the SNR. Later observations (E. A. Helder et al. 2013) have further constrained the distance to $2.5 \pm 0.4$ kpc. The age of an SNR can be estimated from its angular size and expansion velocity. Looking at the size, distance, and expansion speed, the remnant was initially thought to be as old as $10^4$ yr (M. Rosado et al. 1996). However, X-ray observations have provided further insights into the complex dynamics of this SNR. Spectral analysis reveals a strong density contrast between the northeastern and southwestern regions, suggesting different environmental conditions and an off-center supernova explosion within a wind-blown bubble. This scenario accounts for the large size of the SNR—approximately 15 pc in radius—despite its young age. In the southwestern region, earlier X-ray studies have revealed an energy-dependent morphology (J. Vink et al. 1997, 2006; F. Bocchino et al. 2000; K. J. Borkowski et al. 2001). Below 2 keV, the emission is dominated by a thermal component exhibiting an arc-like morphology, whereas emission above 2 keV is better described by a synchrotron component with a high-energy cutoff and low surface brightness (Y. Tsubone et al. 2017). As suggested by S. Broersen et al.

(2014), RCW 86 is much younger than previously thought and possibly compatible with the oldest historical supernova event, recorded in 185 AD (D. H. Clark & F. R. Stephenson 1975; F. R. Stephenson & D. A. Green 2002; J. Vink et al. 2006). This SNR is known to exhibit K-shell emission from low-ionized iron (J. Vink et al. 1997; K. J. Borkowski et al. 2001; M. Ueno et al. 2007; H. Yamaguchi et al. 2011), possibly originating from Type Ia supernova ejecta. The spatial distribution of the Fe K line emission resembles that of the radio emission, but differs from the synchrotron X-ray emission (M. Ueno et al. 2007). This suggests that the physical conditions required to accelerate TeV electrons responsible for synchrotron X-ray production are distinct from those prevailing in regions traced by radio and Fe K line emission.

The many peculiarities of RCW 86 are primarily attributed to its irregular morphology and the small-scale spatial variation of physical parameters across the SNR. This is particularly pronounced in the southwestern region, where Chandra observations have revealed a broad range of expansion velocities, directions, and ambient densities, as well as reflected/reverse shocks sweeping through, possibly indicating a dynamically evolving environment as the shock interacts with the cavity boundary (H. Suzuki et al. 2022).

Space-resolved X-ray polarimetry can be a powerful diagnostic, since the Imaging X-ray Polarimetry Explorer (IXPE) has already observed many young SNRs (J. Vink et al. 2022; R. Ferrazzoli et al. 2023, 2024; P. Zhou et al. 2023, 2025; D. A. Prokhorov et al. 2024) showing differently oriented magnetic fields and polarization degrees (PDs), which seem to be anticorrelated with the ambient density. The southwestern region of RCW 86 thus provides a case study of particular interest for exploring magnetic field topology via X-ray polarimetry within a complex environment shaped by the interaction between the SNR and the wall of a wind-blown bubble. This paper presents the IXPE observations of the southwestern rim of RCW 86, revealing a polarization below the detection sensitivity achieved in 816 ks of total exposure.

---

[56] Deceased

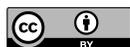

## 2. IXPE Observation and Data Analysis

IXPE is the first X-ray mission entirely dedicated to polarimetry, with the capability to spatially resolve extended sources (M. C. Weisskopf et al. 2022). It is the result of a joint collaboration between the National Aeronautics and Space Administration (NASA) and the Italian Space Agency (ASI). IXPE features three identical but independent telescopes, each equipped with Gas Pixel Detectors (E. Costa et al. 2001; R. Bellazzini et al. 2006; L. Baldini et al. 2021) in detector units (DUs) positioned at the focal plane of X-ray mirrors in Wolter configuration.

IXPE observed the southwestern rim of the RCW 86 SNR during two separate epochs. The first began on 2023 July 23, with a total livetime of 543 ks (ObsID 02001599), and the second began on 2024 January 29, lasting 272 ks (ObsID 02009501), resulting in a combined livetime of about 816 ks. Data reduction and analysis have been carried out with IXPEOBSSIM (v31.1.0, L. Baldini et al. 2022), XSPEC (v12.14.0.h, K. A. Arnaud 1996), and BXA (v4.1.4, J. Buchner et al. 2014). Due to strong background contamination, we developed and applied a new method to characterize the background by extracting (1) the variable component affected by solar flares using time intervals when the Sun illuminated the spacecraft, and (2) the steady component from archival observations.

### 2.1. Background Treatment

IXPE's background is mostly instrumental (A. Di Marco et al. 2023). This means that the largest contribution to the background comes from particles either entering the detector or being converted into X-rays in the detector casing. Even after the rejection of a fraction of events due to particles identified through the shape of the track (A. Di Marco et al. 2023) it still contains at least two residual (X-ray indistinguishable) components: a constant or long-timescale-variable ("static") component and a short-timescale-variable ("flaring") component. Previous efforts to characterize the flaring component relied on sigma-clipping algorithms based on the removal of outlier time bins in the light curve ("deflaring," see F. Xie et al. 2022; F. Marin et al. 2023; P. Zhou et al. 2023; R. Ferrazzoli et al. 2024; P. Kaaret et al. 2024; D. A. Prokhorov et al. 2024; for a nonexhaustive list of past examples). However, a relatively low-rate variable background with a polarized component has been identified in at least one other source, the pulsar wind nebula 3C 58 (N. Bucciantini et al. 2025), even upon deflaring. This indicates the need for more sophisticated background subtraction procedures, especially in samples where the abundance of source counts is insufficient to compensate for the spurious polarization effect introduced by the background (for a more detailed explanation, see Appendix A.1).

*In-Sun/in-eclipse analysis.* There is mounting evidence of the variable background component being azimuthally modulated. This means that there is a larger number of background events that are reconstructed with a preferential direction,[57] thus resulting in a spurious detection of polarization upon reconstruction with IXPE's response functions. As observed by N. Bucciantini et al. (2025), variability has a mild correlation with solar activity and seems to be detected mostly when the Sun is directly illuminating the spacecraft, with a large difference between DUs that seems to arise from their layout with respect to the solar panels and solar boom, which obstruct the direct sunlight. On top of the variable component, IXPE also shows another component that seems to be the same across different DUs and stable at least on timescales of about a year. All the binned products that we analyzed hereafter have been obtained with the following procedure, under the assumption of a uniform background across the DUs[58]:

1. selection of the energy range (see Appendix B);
2. creation of a static background template;
3. creation of binned data products for the background template (polarization cubes,[59] Stokes spectra, polarization maps);
4. creation of in-Sun and in-eclipse data files for the full field of view, i.e., the subset of events acquired when the Sun is illuminating the spacecraft, and when the the spacecraft is in the Earth's shadow;
5. creation of the in-Sun and in-eclipse binned data products;
6. extraction of the flaring component by the subtraction of the in-eclipse data from the in-Sun data;
7. regional selection (polarization cubes or spectra) or spatial binning (polarization maps) and creation of the aforementioned binned data product;
8. two-step subtraction: counts or rates of the flaring component and then of the static component in binned products are subtracted from each channel (Stokes parameters, $I$, $Q$, $U$) of the aforementioned binned data products.

A detailed description and justification of this novel background subtraction approach will be presented in a dedicated methodological paper that is currently in preparation.

### 2.2. World Coordinate System (WCS) Correction

There is a marked spatial misalignment between ObsIDs 02001599 and 02009501, so we performed a manual alignment using a minimization algorithm in the region comprising the three bright features (easily seen in Figure 1) in the western area with $2''$ steps. For each displacement, we fine-binned and smoothed the count map and we calculated

$$\sum_{i,j} \left( \frac{C_{i,j}^{(\text{obs}_2)} - C_{i,j}^{(\text{obs}_1)}}{\sigma_{i,j;12}} \right)^2,$$

where $i$, $j$ are the column and row corresponding to the pixel matrices, $C$ are the counts therein contained for ObsIDs 02009501 (obs$_2$) and 02001599 (obs$_1$) respectively, and $\sigma_{i,j;12}$ is the uncertainty in the counting statistic propagated as the sum of the two statistical samples. The minimum value was obtained for a shift of the central pixel values in the header files of ObsID 02009501 to 288 (TCRPX7) and 326 (TCRPX8), from their initial value of 299, while leaving the celestial coordinates unaltered, thus aligning everything to ObsID 02001599.

---

[57] This is *exactly* how we detect polarization, but caution is exercised because this might be a spurious effect arising from anomalous reconstruction of background events.

[58] This was checked across several observations, but details will follow in a dedicated paper.

[59] Stokes parameters sliced in energy and space.

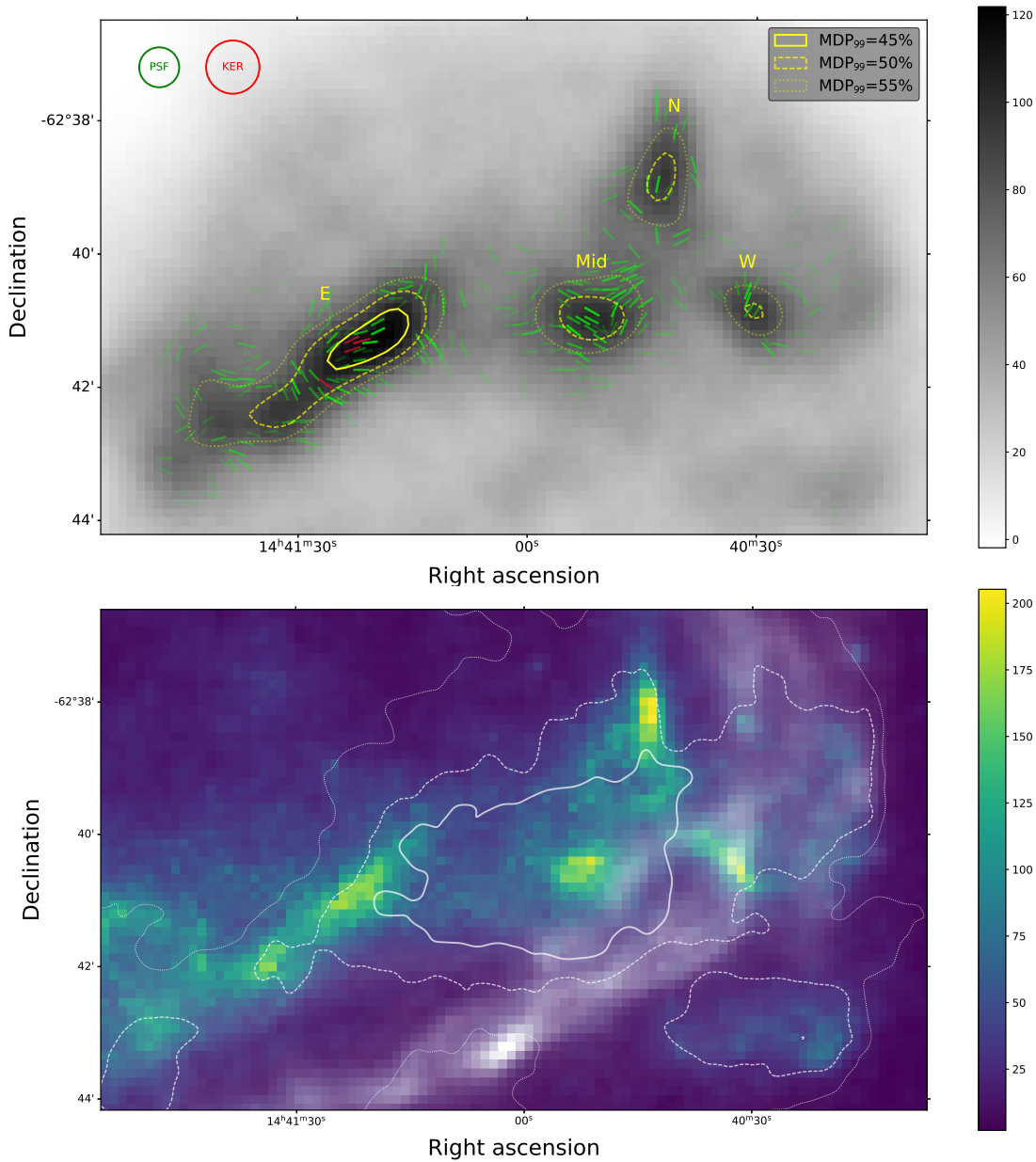

**Figure 1.** Top: background-subtracted, 5 pixel smoothed polarization map of the southwestern rim of RCW 86 in the 2–4.5 keV band. The color map represents the kernel size-integrated Stokes' $I$, while contours represent MDP$_{99}$ levels (from smaller to larger: 45%, 50% 55%), which also serve as upper limits for circular regions centered on each pixel. The vectors represent magnetic field direction, and the color represents significance, with magenta being in excess of $2\sigma$ (the more transparent magenta vectors being at the lower edge of $2\sigma$) and red in excess of $3\sigma$. Despite individual bin fluctuations above $3\sigma$ and a corresponding PD value above the MDP$_{99}$ of the contour encompassing the bin, no significant post-trial detection of polarization is found across the rim. Bottom: XMM-Newton high-energy (2–5 keV) with thermal (0.5–1 keV, white to transparent shading) emission and MeerKAT Stokes $I$ contours overlaid (thicker represent brighter). **Note**: The circular area denoted "KER" represents the kernel size used for smoothing (top hat), whereas the circular area denoted "PSF" represents the half-power diameter of the worst mirror (30″).

## 3. Results

The analysis of RCW 86 has been carried out in the energy range in which the source dominates the static background component, that is, 2–4.5 keV.

### 3.1. Polarization Map

We first searched for a polarized signal by making a polarization map with a fine binning (8″) and smoothing it with a top-hat kernel with a size slightly larger than IXPE's half-power diameter (HPD; 5 pixels, or 40″). The smoothed polarization map is shown in Figure 1. The yellow contours represent the threshold levels for the minimum detectable polarization at the 99% confidence level (MDP$_{99}$), which serve both as our detection criterion and as upper limits for polarization. Since smoothing was applied, each of the pixels of the map is an integration over the kernel used for convolution, and hence represents a circular region of 40″ diameter centered on the pixel. The same applies to the MDP$_{99}$ contours, which encompass regions in which each pixel has that minimum level of MDP$_{99}$ after smoothing of the map.

**Table 1**
Results of the Analysis of the Regions Highlighted in Figure 1

| Region | Polarization Cubes | | | | Spectropolarimetric | | | Chandra Fit Thermal |
|---|---|---|---|---|---|---|---|---|
| | MDP$_{99}$ | Source Counts | Total Counts | $K_{PU}$ | Photon Index | PD$_{max}$ | PD$_{UL}$ | |
| E$_{45}$ | 34% | 2947 | 3420 | 0.232 | 2.48 ± 0.11 | 80% | 31% | 3% |
| E$_{50}$ | 20% | 8783 | 10,477 | 0.081 | 2.54 ± 0.06 | 77% | 19% | ⋯ |
| E$_{55}$ | 15% | 15,695 | 19,284 | 0.186 | 2.53 ± 0.05 | 77% | 19% | ⋯ |
| Mid$_{50}$ | 42% | 2077 | 2486 | 0.181 | 2.72 ± 0.14 | 79% | 32% | 12% |
| Mid$_{55}$ | 27% | 5104 | 6303 | 0.247 | 2.61 ± 0.09 | 80% | 35% | ⋯ |
| N$_{55}$ | 33% | 3223 | 4012 | 0.172 | 2.23 ± 0.11 | 79% | 34% | 20% |
| W$_{55}$ | 46% | 1793 | 2188 | 0.216 | 2.70 ± 0.14 | 79% | 30% | 27% |
| (E + Mid + N)$_{55}$ | 12% | 24,020 | 29,599 | 0.146 | 2.50 ± 0.038 | 79% | 17% | 3% |

**Note.** Regions are marked here with the coordinates (E, Mid, N, W) and the subscript relative to the contours of the MDP$_{99}$ from Figure 1. The last line represents an integrated measurement performed after aligning the polarization vectors to the center of the SNR explosion (R.A. 14$^h$43$^m$, decl. −62°30′, F.-Y. Zhao et al. 2006). Source counts are background-free, and total counts are source and both background components united. The leftmost four columns are relative to the polarization cube analysis and reveal no significant detection, hence MDP$_{99}$ is used for upper limits. The following three columns represent the spectropolarimetric analysis with the Bayes factors of the polarized versus unpolarized model, again revealing no significant detection. In this case, the upper limits coming from the 99% percentile of the posterior distribution for PD are shown. Finally, the rightmost column shows the maximum thermal contribution to the flux estimated from Chandra ObsID 1993 in the regions that we considered (details in the text).

### 3.2. Regional Analysis

We performed a more sensitive search by selecting regions and integrating the full signal therein. In this way, MDP$_{99}$ decreases with the obvious risk of washing out polarization if the magnetic field is not coherent across the full region of integration. We defined regions based on the MDP$_{99}$ contours from Figure 1 and investigated them individually. The smallest regions, providing fewer counts than the convolution kernel alone, have been left out (this includes the one encompassed by the 50% contours in the western blob). The subdivision of these regions is shown in Figure 1 and the results of the polarization cube analysis are summarized in Table 1. Note that wherever we refer to PD in the regional analysis it applies to the polarization of the photon sample, so that the thermal dilution is not accounted for.

None of the analyzed regions have a high enough significance to claim a detection. Being background-subtracted, the reported upper limits are for the synchrotron component under the assumption that the thermal component is negligible. Accounting for dilution by the thermal component can bring the PD upper limit up to about 30% for the Mid region and 39% for the N region, with the E region being essentially unaffected. As done for other SNRs, we attempted to align the Stokes parameters to a presumed center of explosion at R.A. 14$^h$43$^m$, decl. −62°30′ to test whether the radial or circular symmetry is recovered. This has provided higher sensitivity in other extended sources (most notably J. Vink et al. 2022), but no improvement was detected in this case.

Even the most likely value of the PD, even in this case, is far from what is required for a detection (as expected when looking at the polarization map in Figure 1) and only upper limits are reported.

### 3.3. Spectropolarimetric Analysis

We performed a Bayesian spectropolarimetric analysis of the same regions from Section 3.2 using BXA in conjunction with XSPEC. The spectra have been fit with a simple TBabs*polconst*powerlaw model[60] that assumes constant polarization, in which we fixed the TBabs parameter values used by H. Suzuki et al. (2022) for similar zones to $(0.23-0.40) \times 10^{22}$ cm$^{-2}$. We then fixed the polarization to 0, or left it as a free parameter with uniform prior, and compared the evidence of the two models to assess which better described our data with the Bayes factors. Instead of using $p$-values, Bayesian analyses are parameterized based on the evidence of a model Z, which represents the likelihood; by comparing the evidence of different models the most credible can be assessed through Bayes factors $K$ (H. Jeffreys 1939). Since BXA outputs logZ, the Bayes factor is

$$K_{PU} = \exp(\log Z_P - \log Z_U), \quad (1)$$

where the subscripts P and U stand for the polarized and unpolarized models respectively. The Bayes factors of the unpolarized versus polarized models have been calculated and listed in Table 1. For all regions, we also report the measured photon index Γ with its uncertainty, which is used in the following relation to compute the theoretical maximum PD (V. L. Ginzburg & S. I. Syrovatskii 1965):

$$\text{PD}_{\text{MAX}} = \frac{\Gamma}{\Gamma + 2/3}. \quad (2)$$

For all the posterior distributions of polarized models we also calculate the 99% percentile of the PD indicated as PD$_{UL}$ in Table 1, which is in principle an estimation of the same quantity that we measure with MDP$_{99}$ from the polarization cubes. It is important to emphasize two fundamental differences. First, the polarization cube and spectral approaches are different in that spectral analysis accounts for energy dispersion differently. Second, spectral fitting is model-dependent, and the power-law model with constant polarization does not always represent a good model for all of our regions.

---

[60] See the XSPEC model definition.

### 3.4. Thermal Contribution

While the thermal contribution in synchrotron-emitting regions in the 2.0–4.5 keV range is generally considered to be so low that it cannot be fitted, we must also note that our regional analysis includes much broader features as seen through IXPE's HPD. We performed the fits on Chandra data (ObsID 1993) using XSPEC with the model from H. Suzuki et al. (2022) to estimate the contribution of the thermal component to the overall flux using calcFlux. In order to recover the maximum possible thermal component and be able to fit it, we used the largest possible regions corresponding to $MDP_{99} = 55\%$ in Figure 1. The thermal contributions are reported in Table 1, but they are not rigorous because they are based on a Chandra observation and do not completely account for the IXPE's point-spread function. For region E, Chandra coverage is partial, with results based solely on the westernmost part. It is also worth noting that region E, which provides the most constraining upper limit, is the least affected due to its clear separation from the thermally emitting region at the bubble edge (see Figure 1 of B. J. Williams et al. 2011).

## 4. Discussion

The analysis of the southwestern rim of RCW 86 does not reveal any significantly polarized X-ray emission. This has been assessed down to the physical scale (0.2–0.7 pc) at which all SNRs have been studied using IXPE (J. Vink et al. 2022; R. Ferrazzoli et al. 2023, 2024; P. Zhou et al. 2023, 2025; D. A. Prokhorov et al. 2024). Indeed, the map in Figure 1 shows some peaks of polarization that, however, are not coherent across the bright synchrotron structures at the scale of our kernel size of 10 pixels or 40″, corresponding to 0.54 pc at a distance of 2.8 kpc. While appealing, these peaks do not reach the required significance to claim a detection anywhere in the southwestern rim of RCW 86 as all the pixels in the kernel are correlated. Their appearance matches expectations for random fluctuations dominating individual pixels, since all neighboring pixels within the kernel size are correlated and influenced by the same fluctuation. A PD smaller than 50% everywhere on the length scale of 0.54 pc at most indicates a significant departure from a fully coherent magnetic field. A fully coherent field would lead to a theoretical maximum PD of approximately 80%. Integrating the signal over larger domains lowers $MDP_{99}$, but increases the length scale over which we are sampling the coherence of the magnetic field, possibly washing it out in the case of irregularities. The PD upper limits reported in Table 1 indicate the absence of large-scale coherent magnetic field structures across the rim.

The largest finger-shaped region ($E_{55}$) yields our most constraining upper limit of 15% among all individual regions in the southwestern rim of RCW 86. For the smaller parsec-scale regions with higher signal-to-noise ratios ($E_{45}$, $Mid_{50}$, $N_{55}$), we exclude a PD of the synchrotron component exceeding ∼40%. The upper limit obtained by integrating all the regions that roughly obey circular symmetry around the presumed center of the explosion is at the level of 12%.

These values need to be compared with the peak values observed in other SNRs. For Vela Jr. (D. A. Prokhorov et al. 2024), the peak PD was measured to be 85% ± 30%, for RX J1713.7-3946 (R. Ferrazzoli et al. 2024) the peak PD was found to be 46% ± 9%, in SN 1006 (P. Zhou et al. 2025) the turbulence appeared to be very uniform and peak PD was in the range 20%–26%, and for Tycho SNR the peak PD was 23% ± 4%. Finally, the SNR in our sample that has the lowest localized measurement (still not in excess of 5σ) is Cas A (J. Vink et al. 2022; A. Mercuri et al. 2025), which is still at the level of 20%–22%. High PD and magnetic field coherence have previously been associated with low-density environments (P. Slane et al. 2024). The lack of even a single hot spot in excess of 20% PD in the southwestern rim of RCW 86 is unprecedented in our SNR sample, and calls for an explanation about its uniqueness.

The regions labeled in Figure 1 are projected onto the inner parts of the southwestern rim of RCW 86, away from its edge. A lower PD may be an intrinsic characteristic of the rim with an exceptionally complex environment, even by SNR standards. The interaction of the forward shock with dense material can generate reflected shocks that propagate backward into the shocked plasma, as evidenced by the complex shock structures in this region reported by H. Suzuki et al. (2022). Furthermore, the shock–cloud interaction (e.g., H. Sano et al. 2017) amplifies turbulence. The resulting complex magnetic field topology may challenge the recovery of a net polarized signal through spatial integration across the rim.

H. Suzuki et al. (2022) focus on the synchrotron properties of regions Mid, N, and W, studying the shock speed based on measured proper motions. Depending on the physical frame in which the shocks are propagating, different shock speeds can be inferred. For the regions of interest, an interpretation of the proper motion as being associated with the reverse shock propagating into unshocked ejecta yields an extremely high value of ∼10,000 km s$^{-1}$. This high speed of the reverse shock is inferred by H. Suzuki et al. (2022) as the relative velocity between the unshocked ejecta—freely expanding for ≈1840 yr after the supernova explosion—and the shocked material as traced by the proper motion measured in the observer's frame. This represents the maximum shock speed, as more moderate values are recovered if the motion is interpreted as a reflected shock sweeping through the shocked ejecta. More relevant is that the proper motions measured by H. Suzuki et al. (2022) in the regions of interest from Figure 1 appear to be in different directions, which provides an important ingredient for gas-dynamical turbulence, and hence magnetic field turbulence as well, assuming this result is not a projection effect.

The Bohm factor $\eta$, which measures the effectiveness of magnetic turbulence in diffusing electrons across shock fronts, is related to the shock speed and the photon cutoff energy, and can be expressed as follows (V. N. Zirakashvili & F. Aharonian 2007):

$$\eta = 0.4 \text{ keV} E_0^{-1} \left( \frac{v_{sh}}{2000 \text{ km s}^{-1}} \right)^2. \quad (3)$$

Based on IXPE observations of multiple SNRs we also noted a possible correlation between the observed PD and the Bohm factor. This is not surprising, since it is an indicator of turbulence, therefore we expect it to also influence the coherence of the magnetic field and the observed PD. In addition, in X-ray studies, the Bohm factor has also been linked to the magnetic field direction, with higher Bohm factors favoring radially oriented magnetic fields. Below, we discuss the impact of the Bohm factor on polarization in various physical scenarios, in the context of previous IXPE observations of SNRs. If we consider a reverse shock, involving high shock speeds, the shock characteristics become close to those observed in SN 1006, particularly in the JKL region, as reported by P. Zhou et al. (2025), in which a radial magnetic field with a peak value of the PD as high as 30% has been observed. On the other hand, if we use the reflected shock scenario from H. Suzuki et al. (2022) we recover a Bohm factor of $\simeq 1$. The shock characteristics are then closer to those observed in RX J1713.7-3946 and Vela Jr., in which a tangential magnetic field with a PD of at least 35% is found. Importantly, in the reflected shock scenario, the Bohm factor depends on the velocity of a shock relative to the decelerated ejecta. Consequently, for region SW10 (coincident with region Mid), the Bohm factor ranges from 2 to 4 for ejecta that has been decelerated to observer-frame velocities of 1000 to 2000 km s$^{-1}$, and, as noted by H. Suzuki et al. (2022), it drops to $\simeq 1$ for the case of stationary ejecta.

The spatial correlation between the radio synchrotron-emitting region and the Fe K-emitting region suggests that electron acceleration responsible for the radio emission may have occurred in the denser portions of the shocked ejecta. The peak PD value of 15% and the radial magnetic field orientation observed in RCW 86 in the radio band (J. R. Dickel et al. 2001) are compatible with those reported for regions of the forward shock in Tycho's SNR in the X-ray band (R. Ferrazzoli et al. 2023), and correspond to intermediate Bohm factor values. Meanwhile, the anticorrelation between X-ray synchrotron-emitting and Fe K-emitting regions implies distinct electron energy distribution histories. From this perspective, the eastern region warrants focused attention. According to B. J. Williams et al. (2011), the diffuse nonthermal emission observed in this region may result from electrons accelerated much earlier in the evolution of the SNR —up to $\sim 1500$ yr ago—when the shock velocity was very high, prior to its interaction with the bubble wall. While a thin rim of synchrotron emission would have been produced at that time, electrons still capable of producing observable X-rays at the current epoch will have diffused far from the initial acceleration site, yielding an overall extended morphology. If TeV electrons were accelerated early in the remnant's evolution, while the forward shock was still propagating through the low-density bubble, a high PD would be expected due to the associated low density. The present X-ray polarimetric analysis challenges scenarios that invoke physical conditions leading to a high PD.

## 5. Summary

The analysis of faint extended and morphologically complex structures remains one of the most challenging tasks for IXPE. As such, an advanced background characterization is mandatory to extract the polarization information from faint dilute sources such as SNRs. The background characterization of IXPE has proven to be mature enough to extract meaningful information despite background variability. This is especially critical for observations carried out close to the solar maximum. The upper limits provided for the PD exclude the presence of a coherent magnetic field not only when integrated throughout the southwestern rim of RCW 86, but in every single synchrotron-bright feature hereby contained and at all scales that we have probed. In the eastern region, upper limits at the level of 15% are above the levels detected in previous studies of SNRs. However, the same limits are not as stringent if we extend the search to other, smaller regions, for which the upper limit rises to about 30%–40%. This remains somewhat surprising, considering that we are approaching the expected maximum polarization level for hot spots, and still falls well below the values found in sources such as RX J1713.7-3946 or Vela Jr.

The derived limits challenge the possibility of synchrotron X-ray production by reflected shocks when shock speeds are as low as those measured in the observer's frame. Such slow shocks would have given rise to polarization properties close to those observed for RX J1713.7-3946 and Vela Jr. (tangential B field, strong PD). If the reverse shock model is to be considered, the Bohm factors derived from spectral analysis (H. Suzuki et al. 2022), combined with the assumption of a free-expansion phase, may result in a radial magnetic field with a PD close to our upper limits. Another possibility is that the shocks producing X-ray synchrotron emission propagate through tenuous portions of the shocked ejecta in the southwestern region of RCW 86. In this frame, they yield intermediate Bohm factors that resemble the situation in Tycho and Cas A SNRs, which display relatively low PDs and radial magnetic fields.

## Acknowledgments

The Imaging X-ray Polarimetry Explorer (IXPE) is a joint US and Italian mission. The US contribution is supported by the National Aeronautics and Space Administration (NASA) and led and managed by its Marshall Space Flight Center (MSFC), with industry partner Ball Aerospace (contract NNM15AA18C). The Italian contribution is supported by the Italian Space Agency (Agenzia Spaziale Italiana, ASI) through contract ASI-OHBI-2022-13-I.0, agreements ASI-INAF-2022-19-HH.0 and ASI-INFN-2017.13-H0, and its Space Science Data Center (SSDC) with agreements ASI-INAF-2022-14-HH.0 and ASI-INFN 2021-43-HH.0, and by the Istituto Nazionale di Astrofisica (INAF) and the Istituto Nazionale di Fisica Nucleare (INFN) in Italy. This research used data products provided by the IXPE Team (MSFC, SSDC, INAF, and INFN) and distributed with additional software tools by the High-Energy Astrophysics Science Archive Research Center (HEASARC), at NASA Goddard Space Flight Center (GSFC). R.F., is partially supported by MAECI with grant CN24GR08 "GRBAXP: Guangxi–Rome Bilateral Agreement for X-ray Polarimetry in Astrophysics." P.Z. acknowledges the support from NSFC grant No. 12273010. F.X. is supported by National Key R&D Program of China (grant No. 2023YFE0117200), and National Natural Science Foundation of China (grant Nos. 12373041 and 12422306), and Bagui Scholars Program (XF). C.-Y.N. is supported by a GRF grant of the Hong Kong Government under HKU 17304524. I.L was funded by the European Union ERC-2022-STG – BOOTES – 101076343. Views and opinions expressed are however those of the author(s) only and do not necessarily reflect those of the European Union or the European Research Council Executive Agency. Neither the European Union nor the granting authority can be held responsible for them.

## Appendix A
## Notes on Nonstandard Analysis

The anomalous background characteristics of ObsID 02001599 required the application of nonstandard analysis techniques. While the details regarding both the nature of the background and how to address it will be presented in an upcoming publication (S. Silvestri et al. 2026, in preparation), here we briefly summarize the methods used to address it in this particular observation.

### A.1. Handling the Background

The background of IXPE is composed of an easily identifiable particle component (A. Di Marco et al. 2023) and a residual component that can be further subcategorized into a static X-ray component of instrumental origin and a variable X-ray component correlated with solar activity that can result in spurious polarization when analyzed with our response functions. This last component has historically been addressed through deflaring, but since we do not have a strong detection of polarization, we decided to opt for a more sophisticated solution for RCW 86.

*Deflaring RCW 86*. The deflaring of RCW 86 through standard sigma-clipping algorithms proved to be inefficient. Figure 2 shows that, relative to a pure background sample taken roughly contemporaneously with the first-epoch observation, a persistent polarization remains in the full dataset even after radical deflaring.

It can be further seen that the flares happen mostly when the Sun is shining on the detector. By using the housekeeping data and the dedicated columns indicating the seconds remaining in eclipse, we created a mask to separate the in-Sun and in-eclipse events and overlaid it on the light curve of RCW 86. An effect of the Sun shining on the spacecraft is clearly visible in Figure 3, where most of the time intervals affected by severe flaring seem to be typical of the in-Sun phases, and this can be used for flare characterization, instead of subtraction. This seems to be evident also in Figure 4, where the typical spectral signature of the flares (peak at ∼1.5 keV) can be seen only in the in-Sun data sample.

The procedure described in Section 2.1 is based on the creation of these datasets, the reconstruction of the livetime, and the subtraction of charged particles, and as it can be seen in Table 2, this method recovers the $\sqrt{t}$ statistics.

**Table 2**
Comparison Between the Upper Limits at 99% Confidence Level Recovered from the In-eclipse Data Only and the Full Dataset with Subtraction

| Region | $MDP_{99}$ (Full Dataset) | $MDP_{99}$ (In-eclipse) |
| --- | --- | --- |
| $E_{45}$ | 34% | 52% |
| $E_{50}$ | 20% | 31% |
| $E_{55}$ | 15% | 24% |
| $Mid_{50}$ | 42% | 66% |
| $Mid_{55}$ | 27% | 43% |
| $N_{55}$ | 33% | 50% |
| $W_{55}$ | 46% | 73% |
| $(E + Mid + N)_{55}$ | 12% | 19% |

**Note.** The livetime for the in-eclipse data is 334 ks compared to the full livetime of 816 ks.

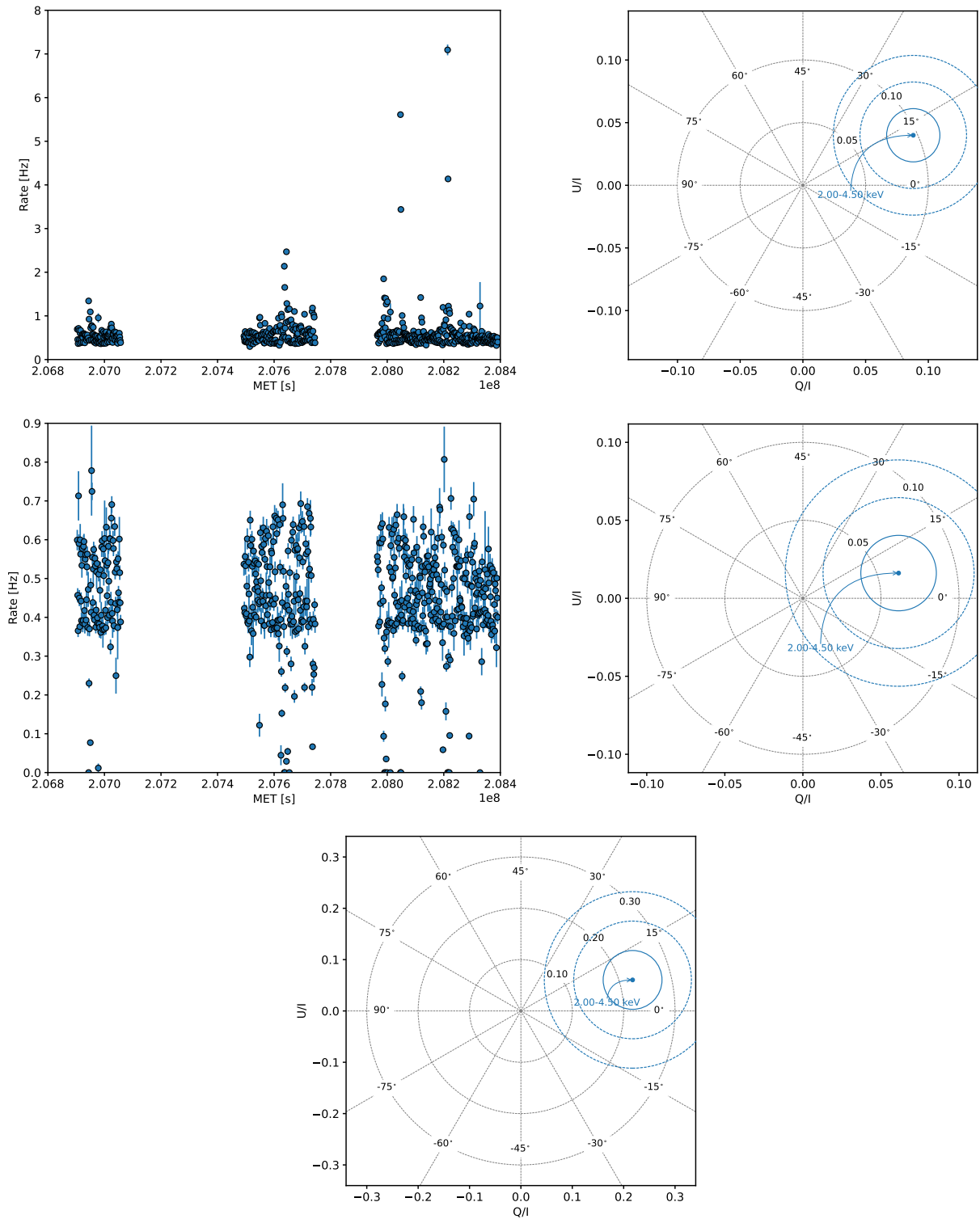

**Figure 2.** Effect of sigma clipping on polarization. On the top, the light curve (left) of the first-epoch on-source dataset shows evident variability as well as significant overall polarization (right). Middle plots show the effect of a radical deflaring (25% quantile clipping) on the light curve as well as a residual polarization. The bottom plot shows the polarization of the background only (e.g., source occulted by the Earth), consistent with the angles of the background before and after deflaring.

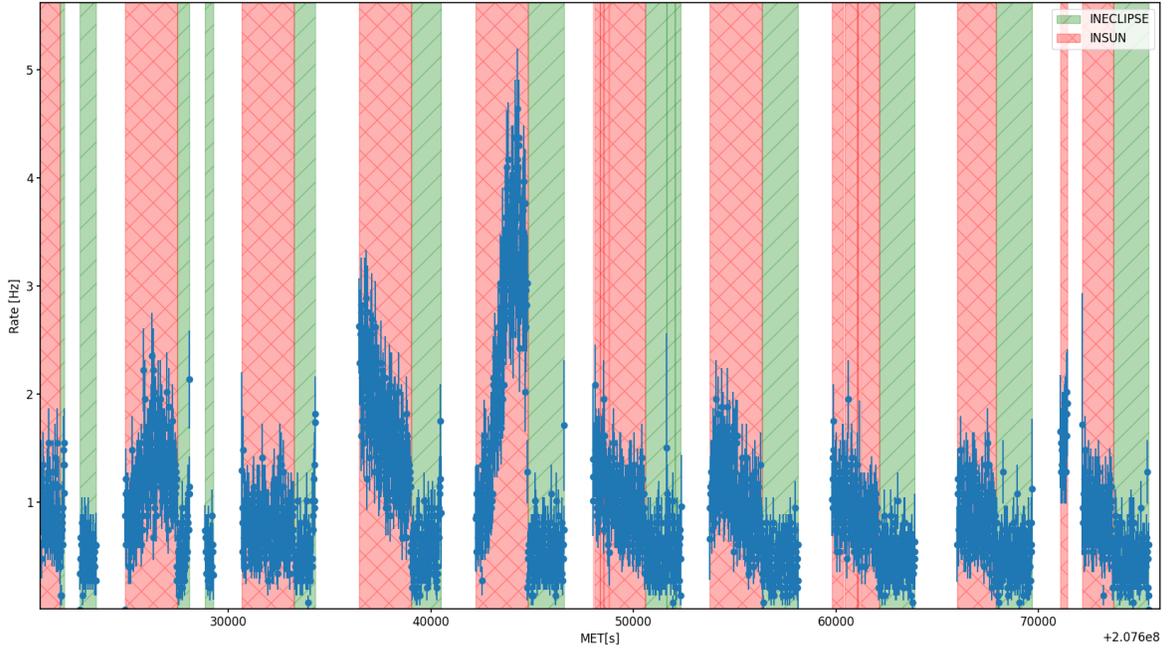

**Figure 3.** Details of the light curve from DU2 in the first epoch of the observation. It can be seen that most of the high-rate events pile up when the Sun shines on the spacecraft ("in-Sun", red or double-hatched area), while the light curve during the eclipse (green or single-hatched) seems more stable. Note: white bands represent bad time intervals (i.e., occultations, South Atlantic Anomaly (SAA), anomalies). The vertical lines with stronger coloring are artifacts arising from short bad time intervals and edge-coloring in the plotting. Bins close to the edge of white areas can have a slightly increased rate due to the vicinity of the SAA.

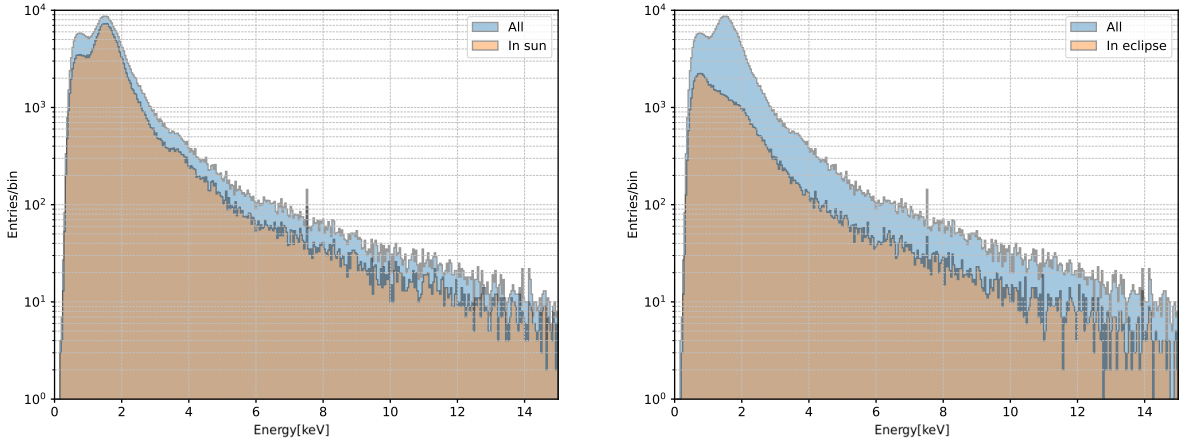

**Figure 4.** Difference between spectra obtained during the in-Sun phase (left) and the in-eclipse phase (right) and the overall spectrum for the case of DU2 and the first segment of our observation (ObsID 02001599). The prominent feature at ∼ 1.5 keV already associated with flaring activity completely disappears if we exclude the in-Sun data.

## Appendix B
## Assessing the Ideal Energy Range for Polarimetry

In order to assess the ideal energy range for our analysis we performed a mosaic simulation similar to the one performed by J. Vink et al. (2022). Mosaic simulations have the goal of assessing the regions and energy ranges in which the synchrotron component, which carries polarimetric information, is dominant. By using Chandra ObsID 1993 we segmented the field of view with a 12 x 12 square grid, and for each segment we subtracted a blank sky background and modeled the spectrum as a power law (synchrotron) and a postshock (thermal) component (*TBabs*<sup>*</sup>*(powerlaw+vps)*) like H. Suzuki et al. (2022). We then proceeded to simulate the observation with the CHANDRA2IXPE utility from IXPEOBSSIM. For every box we then reproduced the spatial distribution of the total count rates and convolved them with the IXPE response function. On top of all the spectral components we added our static background model (bkg). The result is a simulated observation from which we can select the same region and extract the spectrum. Figure 5 shows an example of the resulting simulated spectrum for region Mid, the one in which the greatest ratio of power-law to background flux is recovered. It can be seen that the thermal component is always at lower energies and the synchrotron component dominates the background only below 4.5 keV.

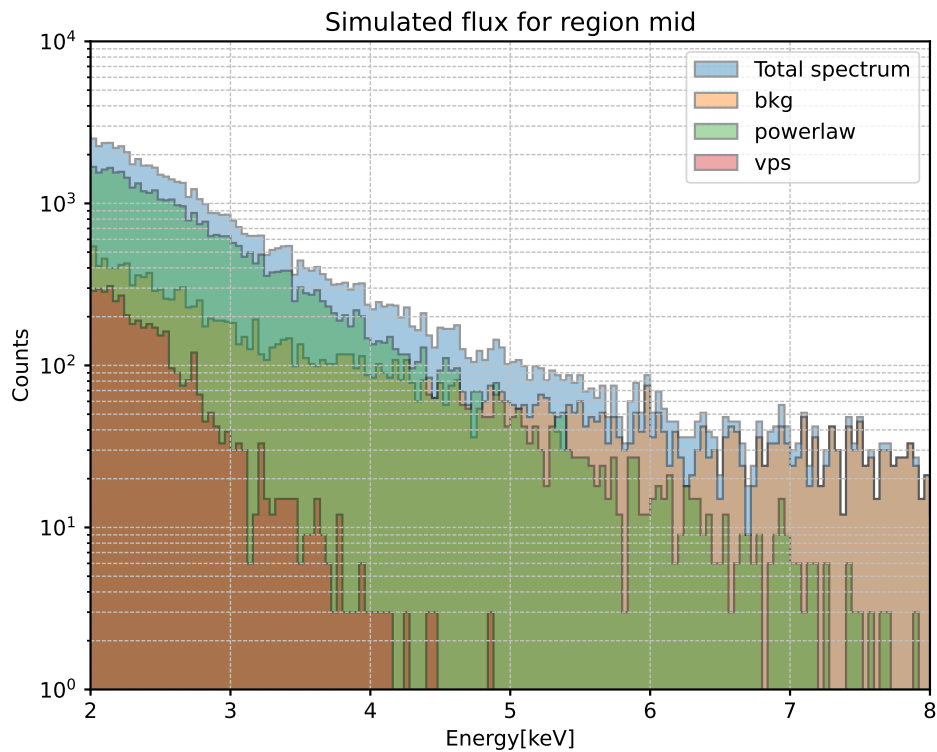

**Figure 5.** Spectral components in the Mid section from a simulated observation in the full 2–8 keV energy range of IXPE. The power-law component, relevant for polarization, shines above the background up to 4.5 keV. Thermal components—in all regions (only region mid is shown here )—are always subdominant and affect the low-energy part of the spectrum, so that the only relevant quantity is the ratio of power-law to background (bkg) flux.

**ORCID iDs**

Stefano Silvestri ⓘ https://orcid.org/0000-0002-8665-0105
Dmitry Prokhorov ⓘ https://orcid.org/0000-0001-6511-4330
Jacco Vink ⓘ https://orcid.org/0000-0002-4708-4219
Patrick Slane ⓘ https://orcid.org/0000-0002-6986-6756
Yi-Jung Yang ⓘ https://orcid.org/0000-0001-9108-573X
Niccolò Bucciantini ⓘ https://orcid.org/0000-0002-8848-1392
Riccardo Ferrazzoli ⓘ https://orcid.org/0000-0003-1074-8605
Ping Zhou ⓘ https://orcid.org/0000-0002-5683-822X
Enrico Costa ⓘ https://orcid.org/0000-0003-4925-8523
Nicola Omodei ⓘ https://orcid.org/0000-0002-5448-7577
Chi-Yung Ng ⓘ https://orcid.org/0000-0002-5847-2612
Paolo Soffitta ⓘ https://orcid.org/0000-0002-7781-4104
Martin C. Weisskopf ⓘ https://orcid.org/0000-0002-5270-4240
Luca Baldini ⓘ https://orcid.org/0000-0002-9785-7726
Alessandro Di Marco ⓘ https://orcid.org/0000-0003-0331-3259
Victor Doroshenko ⓘ https://orcid.org/0000-0001-8162-1105
Jeremy Heyl ⓘ https://orcid.org/0000-0001-9739-367X
Philip Kaaret ⓘ https://orcid.org/0000-0002-3638-0637
Frédéric Marin ⓘ https://orcid.org/0000-0003-4952-0835
Tsunefumi Mizuno ⓘ https://orcid.org/0000-0001-7263-0296
Melissa Pesce-Rollins ⓘ https://orcid.org/0000-0003-1790-8018
Carmelo Sgrò ⓘ https://orcid.org/0000-0001-5676-6214
Douglas A. Swartz ⓘ https://orcid.org/0000-0002-2954-4461
Toru Tamagawa ⓘ https://orcid.org/0000-0002-8801-6263
Fei Xie ⓘ https://orcid.org/0000-0002-0105-5826
Iván Agudo ⓘ https://orcid.org/0000-0002-3777-6182
Lucio A. Antonelli ⓘ https://orcid.org/0000-0002-5037-9034
Matteo Bachetti ⓘ https://orcid.org/0000-0002-4576-9337
Wayne H. Baumgartner ⓘ https://orcid.org/0000-0002-5106-0463
Ronaldo Bellazzini ⓘ https://orcid.org/0000-0002-2469-7063
Stefano Bianchi ⓘ https://orcid.org/0000-0002-4622-4240
Stephen D. Bongiorno ⓘ https://orcid.org/0000-0002-0901-2097
Raffaella Bonino ⓘ https://orcid.org/0000-0002-4264-1215
Alessandro Brez ⓘ https://orcid.org/0000-0002-9460-1821
Fiamma Capitanio ⓘ https://orcid.org/0000-0002-6384-3027
Simone Castellano ⓘ https://orcid.org/0000-0003-1111-4292
Elisabetta Cavazzuti ⓘ https://orcid.org/0000-0001-7150-9638
Chien-Ting Chen ⓘ https://orcid.org/0000-0002-4945-5079
Stefano Ciprini ⓘ https://orcid.org/0000-0002-0712-2479
Alessandra De Rosa ⓘ https://orcid.org/0000-0001-5668-6863
Ettore Del Monte ⓘ https://orcid.org/0000-0002-3013-6334
Laura Di Gesu ⓘ https://orcid.org/0000-0002-5614-5028
Niccolò Di Lalla ⓘ https://orcid.org/0000-0002-7574-1298
Immacolata Donnarumma ⓘ https://orcid.org/0000-0002-4700-4549
Michal Dovčiak ⓘ https://orcid.org/0000-0003-0079-1239
Steven R. Ehlert ⓘ https://orcid.org/0000-0003-4420-2838
Teruaki Enoto ⓘ https://orcid.org/0000-0003-1244-3100
Yuri Evangelista ⓘ https://orcid.org/0000-0001-6096-6710
Sergio Fabiani ⓘ https://orcid.org/0000-0003-1533-0283
Javier A. García ⓘ https://orcid.org/0000-0003-3828-2448
Shuichi Gunji ⓘ https://orcid.org/0000-0002-5881-2445
Wataru Iwakiri ⓘ https://orcid.org/0000-0002-0207-9010
Svetlana G. Jorstad ⓘ https://orcid.org/0000-0001-6158-1708
Vladimir Karas ⓘ https://orcid.org/0000-0002-5760-0459
Fabian Kislat ⓘ https://orcid.org/0000-0001-7477-0380
Jeffery J. Kolodziejczak ⓘ https://orcid.org/0000-0002-0110-6136
Henric Krawczynski ⓘ https://orcid.org/0000-0002-1084-6507
Fabio La Monaca ⓘ https://orcid.org/0000-0001-8916-4156

Luca Latronico https://orcid.org/0000-0002-0984-1856
Ioannis Liodakis https://orcid.org/0000-0001-9200-4006
Simone Maldera https://orcid.org/0000-0002-0698-4421
Alberto Manfreda https://orcid.org/0000-0002-0998-4953
Andrea Marinucci https://orcid.org/0000-0002-2055-4946
Alan P. Marscher https://orcid.org/0000-0001-7396-3332
Herman L. Marshall https://orcid.org/0000-0002-6492-1293
Francesco Massaro https://orcid.org/0000-0002-1704-9850
Giorgio Matt https://orcid.org/0000-0002-2152-0916
Ikuyuki Mitsuishi https://orcid.org/0000-0002-9901-233X
Fabio Muleri https://orcid.org/0000-0003-3331-3794
Michela Negro https://orcid.org/0000-0002-6548-5622
Stephen L. O'Dell https://orcid.org/0000-0002-1868-8056
Chiara Oppedisano https://orcid.org/0000-0001-6194-4601
Alessandro Papitto https://orcid.org/0000-0001-6289-7413
George G. Pavlov https://orcid.org/0000-0002-7481-5259
Abel L. Peirson https://orcid.org/0000-0001-6292-1911
Matteo Perri https://orcid.org/0000-0003-3613-4409
Pierre-Olivier Petrucci https://orcid.org/0000-0001-6061-3480
Maura Pilia https://orcid.org/0000-0001-7397-8091
Andrea Possenti https://orcid.org/0000-0001-5902-3731
Juri Poutanen https://orcid.org/0000-0002-0983-0049
Simonetta Puccetti https://orcid.org/0000-0002-2734-7835
Brian D. Ramsey https://orcid.org/0000-0003-1548-1524
John Rankin https://orcid.org/0000-0002-9774-0560
Ajay Ratheesh https://orcid.org/0000-0003-0411-4243
Oliver J. Roberts https://orcid.org/0000-0002-7150-9061
Roger W. Romani https://orcid.org/0000-0001-6711-3286
Gloria Spandre https://orcid.org/0000-0003-0802-3453
Fabrizio Tavecchio https://orcid.org/0000-0003-0256-0995
Roberto Taverna https://orcid.org/0000-0002-1768-618X
Allyn F. Tennant https://orcid.org/0000-0002-9443-6774
Nicholas E. Thomas https://orcid.org/0000-0003-0411-4606
Francesco Tombesi https://orcid.org/0000-0002-6562-8654
Alessio Trois https://orcid.org/0000-0002-3180-6002
Sergey S. Tsygankov https://orcid.org/0000-0002-9679-0793
Roberto Turolla https://orcid.org/0000-0003-3977-8760
Kinwah Wu https://orcid.org/0000-0002-7568-8765
Silvia Zane https://orcid.org/0000-0001-5326-880X